\documentclass[aps,prl,twocolumn,superscriptaddress,showpacs]{revtex4-1}
\usepackage{graphicx,amsmath,amssymb,bm}

\usepackage[usenames]{color}

\begin{document} 
 
\title{Quantum Monte Carlo Calculations with 
Chiral Effective Field Theory Interactions}

\author{A.\ Gezerlis}
\email[E-mail:~]{gezerlis@theorie.ikp.physik.tu-darmstadt.de}
\affiliation{Institut f\"ur Kernphysik,
Technische Universit\"at Darmstadt, 64289 Darmstadt, Germany}
\affiliation{ExtreMe Matter Institute EMMI,
GSI Helmholtzzentrum f\"ur Schwerionenforschung GmbH, 64291 Darmstadt, Germany}
\author{I.\ Tews}
\affiliation{Institut f\"ur Kernphysik,
Technische Universit\"at Darmstadt, 64289 Darmstadt, Germany}
\affiliation{ExtreMe Matter Institute EMMI,
GSI Helmholtzzentrum f\"ur Schwerionenforschung GmbH, 64291 Darmstadt, Germany}
\author{E.\ Epelbaum} 
\affiliation{Institut f\"ur Theoretische Physik II, Ruhr-Universit\"at Bochum, 
44780 Bochum, Germany}
\author{S.\ Gandolfi}
\affiliation{Theoretical Division, Los Alamos National Laboratory,
Los Alamos, NM 87545, USA}
\author{K.\ Hebeler}
\affiliation{Department of Physics, The Ohio State University, 
Columbus, OH 43210, USA}
\author{A.\ Nogga}
\affiliation{Institut f\"ur Kernphysik, Institute for Advanced Simulation
and J\"ulich Center for Hadron Physics,
Forschungszentrum J\"ulich, 52425 J\"ulich, Germany}
\author{A.\ Schwenk}
\affiliation{ExtreMe Matter Institute EMMI,
GSI Helmholtzzentrum f\"ur Schwerionenforschung GmbH, 64291 Darmstadt, Germany}
\affiliation{Institut f\"ur Kernphysik,
Technische Universit\"at Darmstadt, 64289 Darmstadt, Germany}

\begin{abstract} 
We present the first quantum Monte Carlo (QMC) calculations with
chiral effective field theory (EFT) interactions. To achieve this, we
remove all sources of nonlocality, which hamper the inclusion in QMC calculations, in nuclear forces to next-to-next-to-leading order. We perform auxiliary-field diffusion Monte Carlo (AFDMC) calculations for
the neutron matter energy up to saturation density based on local
leading-order, next-to-leading order, and next-to-next-to-leading order nucleon-nucleon interactions. Our results exhibit a systematic order-by-order convergence in chiral EFT and provide nonperturbative benchmarks with theoretical uncertainties. For the softer interactions, perturbative calculations are in excellent agreement with the AFDMC results. This work paves the way for QMC calculations with systematic chiral EFT interactions for nuclei and nuclear matter, for testing the
perturbativeness of different orders, and allows for matching to
lattice QCD results by varying the pion mass.
\end{abstract} 

 
\maketitle 

Chiral effective field theory (EFT) has revolutionized the theory of nuclear forces by 
providing a systematic expansion for strong interactions at low
energies based on the symmetries of quantum
chromodynamics~\cite{Epelbaum:2009a,Entem:2011,Hammer:2013}.
Chiral interactions have been successfully employed in calculations
of the structure and reactions of light nuclei~\cite{fewbody,NCSM1,%
NCSM2,nuclattice}, medium-mass nuclei~\cite{SM1,SM2,CC1,CC2,IMSRG,%
Gorkov,CaNature}, and nucleonic matter~\cite{Munich,NLOlattice,Hebeler:2010a,%
nucmatt,Hebeler:2010b,Tews:2013,Holt2:2012}. While continuum quantum Monte Carlo (QMC) methods are
very precise for strongly interacting systems~\cite{Ceperley,PTEP},
including neutron matter~\cite{Gezerlis:2008,Magierski:2011,Gezerlis:2011,%
Gandolfi:2012,Gezerlis:2012}, and have provided pioneering calculations
of light nuclei~\cite{Pudliner:1997,Pieper:2008}, QMC methods have not been 
used with chiral EFT interactions due to nonlocalities in their
present implementation in momentum space. In this Letter, we take up
this challenge and combine the accuracy of QMC methods with the
systematic chiral EFT expansion. As an application, we study the
neutron matter equation of state up to nuclear densities. Neutron matter
constitutes an exciting system because of its connections to ultracold
atoms and its importance for the physics of neutron-rich nuclei,
neutron stars, and supernovae. Our work opens up nonperturbative
benchmarks of nuclear matter for astrophysics, including studies of
hyperons, based on chiral EFT, as well as the matching to the underlying
theory of QCD through lattice simulations.

First, we explain how to remove all sources of nonlocality in chiral
EFT interactions to next-to-next-to-leading order (N$^2$LO) and present local nucleon-nucleon (NN)
interactions at leading-order (LO), next-to-leading order (NLO), and
N$^2$LO based on Ref.~\cite{Freunek:2007}. We use the developed chiral
potentials for the first time in QMC calculations to study neutron
matter order by order including theoretical uncertainties. The
nonperturbative QMC results provide many-body benchmarks and enable us
to test perturbative calculations for the same interactions.

The difficulty of handling nonlocal interactions in QMC methods
(see also Ref.~\cite{Lynn:2012}) results from how interactions enter. 
Continuum QMC methods are based
on a path-integral evaluation using propagators of the form:
\begin{equation}
G({\bf R},{\bf R}^{\prime}; \delta \tau) = \langle{\bf R}|e^{-\delta \tau \, \widehat{O}}|
{\bf R}^{\prime}\rangle \,,
\label{eq:prop}
\end{equation}
where ${\bf R} = ({\bf r}_{1},{\bf r}_{2} \ldots {\bf r}_{N})$ is the
configuration vector of all $N$ particles (plus spins
and other quantum numbers), $\delta \tau$ is a step in the
imaginary-time evolution, and the operator $\widehat{O}$ takes into
account the kinetic energy and the interaction part of the
Hamiltonian. The implementation of continuum QMC methods relies on
being able to separate all momentum dependences as a quadratic
$\sum_{i=1}^N {\bf p}_i^2$ term, which is the case for local
interactions, but not for general momentum-dependent, nonlocal
interactions (spin-orbit interactions, linear in momentum, are
manageable). In the local case, the propagator for the
momentum-dependent part is a Gaussian integral
that can be evaluated analytically, and the effects of interactions concern only the positions of the particles.

Chiral EFT interactions are based on a momentum expansion and are
therefore naturally formulated in momentum space~\cite{Epelbaum:2009a,%
Entem:2011}. To regularize interactions at high momenta, one employs
regulator functions, usually of the form $f(p) =
e^{-(p/\Lambda)^{2n}}$ and $f(p')$, where ${\bf p} = ({\bf p}_1 -
{\bf p}_2)/2$ and ${\bf p}' = ({\bf p}'_1 - {\bf p}'_2)/2$ are the
incoming and outgoing relative momenta, respectively. Upon Fourier
transformation, this leads to nonlocal interactions $V({\bf r}, {\bf r}')$
already due to the choice of regulator functions. The other sources of
nonlocality in chiral EFT are due to contact interactions that depend
on the momentum transfer in the exchange channel ${\bf k} = ({\bf p'}
+{\bf p})/2$ and to ${\bf k}$-dependent parts in pion-exchange
contributions beyond N$^2$LO. In contrast, dependences on the momentum
transfer ${\bf q} = {\bf p'} -{\bf p}$ are local, and lead to nonlocalities
only because of the regulator functions used.

To avoid regulator-generated nonlocalities for the long-range
pion-exchange parts of chiral EFT interactions, we use the local
coordinate-space expressions for the LO one-pion-exchange and NLO and
N$^2$LO two-pion-exchange interactions~\cite{DR,SF} and regulate them
directly in coordinate space using the function $f_{\rm long}(r) = 1 -
e^{-(r/R_0)^4}$, which smoothly cuts off interactions at short
distances $r < R_0$ while leaving the long-range parts unchanged. So,
$R_0$ takes over the role of the cutoff $\Lambda$ in momentum
space. To regularize the pion loop integrals of the two-pion-exchange
contributions, we use a spectral-function regularization~\cite{SF}
with cutoff $\widetilde{\Lambda}=800 \, {\rm MeV}$. For the N$^2$LO
two-pion-exchange interactions we take the low-energy constants
$c_1=-0.81 \, {\rm GeV}^{-1}, c_3=-3.4 \, {\rm GeV}^{-1}$, and
$c_4=3.4 \, {\rm GeV}^{-1}$ as in the momentum-space N$^2$LO
potential of Ref.~\cite{EGMN2LO}.

To remove the ${\bf k}$-dependent contact interactions to N$^2$LO, we
make use of the freedom to choose a basis of short-range operators in
chiral EFT interactions (similar to Fierz ambiguities). At LO, one
usually considers the two momentum-independent contact interactions
$C_S + C_T \, {\bm \sigma}_1 \cdot {\bm \sigma}_2$. However, it is
equivalent to choose any two of the four operators $\openone, \, {\bm
\sigma}_1 \cdot {\bm \sigma}_2, \, {\bm \tau}_1 \cdot {\bm \tau}_2$,
and ${\bm \sigma}_1 \cdot {\bm \sigma}_2 \, {\bm \tau}_1 \cdot {\bm
\tau}_2$, with spin and isospin operators ${\bm \sigma}_i, {\bm
\tau}_i$, because there are only two S-wave channels due to the
Pauli principle. It is a convention in present chiral EFT interactions
to neglect the isospin dependence, which is then generated from the
exchange terms.

We use this freedom to keep at NLO (order $Q^2$) an isospin-dependent
$q^2$ contact interaction and an isospin-dependent $({\bm \sigma}_1
\cdot {\bf q})({\bm \sigma}_2 \cdot {\bf q})$ tensor part in favor of
a nonlocal $k^2$ contact interaction and a nonlocal $({\bm \sigma}_1
\cdot {\bf k})({\bm \sigma}_2 \cdot {\bf k})$ tensor part. This leads
to the following seven linearly independent contact interactions at
NLO that are local,
\begin{align}
V^{\rm NLO}_{\rm short} &= C_1 \, q^2 + C_2 \, q^2 \, 
{\bm \tau}_1 \cdot {\bm \tau}_2 \nonumber \\
&+ \bigl(C_3 \, q^2 + C_4 \, q^2 \, {\bm \tau}_1 \cdot {\bm \tau}_2 \bigr)
\, {\bm \sigma}_1 \cdot {\bm \sigma}_2 \nonumber \\
&+ i \, \frac{C_5}{2} \, ({\bm \sigma}_1 + {\bm \sigma}_2) \cdot
{\bf q} \times {\bf k} \nonumber \\
&+ C_6 \, ({\bm \sigma}_1 \cdot {\bf q})({\bm \sigma}_2 \cdot {\bf q}) 
\nonumber \\
&+ C_7 \, ({\bm \sigma}_1 \cdot {\bf q})({\bm \sigma}_2 \cdot {\bf q}) 
\, {\bm \tau}_1 \cdot {\bm \tau}_2 \,,
\label{eq:NLOshort}
\end{align}
where the only ${\bf k}$-dependent contact interaction ($C_5$) is a
spin-orbit potential. Because at NLO the only two possible momentum
operators allowed by symmetries are $q^2$ and $k^2$ (or equivalently
$p^2+p'^2$ and ${\bf p} \cdot {\bf p}'$), and similarly for the tensor
parts, it is thus possible to remove all sources of nonlocality in
chiral EFT to N$^2$LO. In addition, the leading 3N forces at N$^2$LO
can be constructed as local interactions~\cite{Petr,Lovato}, but we
will first focus on QMC calculations with chiral NN interactions. The
next-higher order ($Q^4$) NN contact interactions enter at N$^3$LO,
and there are too many possible operators involving ${\bf k}$, so that
they cannot be traded for isospin dependence completely. Therefore,
chiral EFT interactions will contain nonlocal terms at N$^3$LO, but
one may expect that these high-order nonlocal parts can be treated
perturbatively.

\begin{table}[t]
\begin{center}
\caption{Short-range couplings for $R_0=1.2 \, {\rm fm}$ at LO, NLO,
and N$^2$LO (with a spectral-function cutoff $\widetilde{\Lambda}=800 \,
{\rm MeV}$)~\cite{Freunek:2007}. The couplings $C_{1-7}$ are given in 
fm$^4$ while the rest are in fm$^2$.\label{tab:contacts}}
\begin{tabular}{lccc}
\hline
& LO & NLO & N$^2$LO \\
\hline
$C_S$ & $-1.83406$ & $-0.64687$ & \,\,\,\,\,$1.09225$ \\
$C_T$ & \,\,\,\,\,$0.15766$ & \,\,\,\,\,$0.58128$ & \,\,\,\,\,$0.24388$ \\
$C_1$ & & \,\,\,\,\,$0.18389$ & $-0.13784$ \\
$C_2$ & & \,\,\,\,\,$0.15591$ & \,\,\,\,\,$0.07001$ \\
$C_3$ & & $-0.13768$ & $-0.13017$ \\
$C_4$ & & \,\,\,\,\,$0.02811$ & \,\,\,\,\,$0.02089$ \\
$C_5$ & & $-1.99301$ & $-1.82601$ \\
$C_6$ & & \,\,\,\,\,$0.26774$ & \,\,\,\,\,$0.18700$ \\
$C_7$ & & $-0.25784$ & $-0.24740$ \\
$C_{nn}$ & & & \,\,\,\,\,$0.05009$ \\
\hline
\end{tabular}
\end{center}
\end{table}

\begin{figure*}[t]
\begin{center}
\includegraphics[width=\textwidth,clip=]{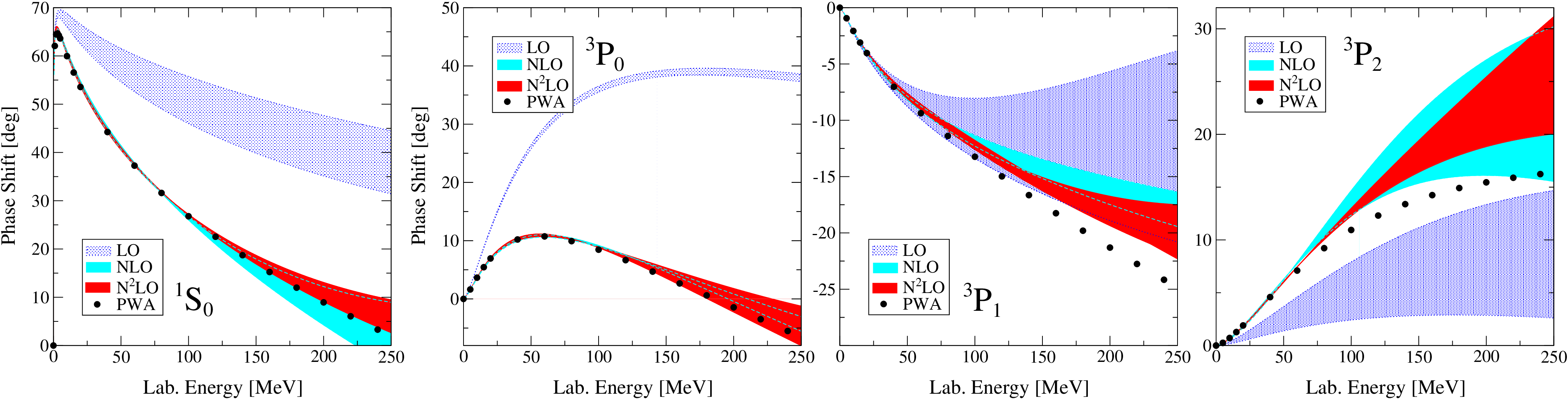}
\caption{(Color online) Neutron-proton phase shifts as a function of 
laboratory energy $E_{\rm lab} = 2 p^2/m$ 
in the $^1$S$_0$, $^3$P$_0$, $^3$P$_1$, and
$^3$P$_2$ partial waves (from left to right) in comparison to the
Nijmegen partial-wave analysis (PWA)~\cite{Stoks:1993}. The LO, NLO,
and N$^2$LO bands are obtained by varying $R_0$ between $0.8-1.2 \,
{\rm fm}$ (with a spectral-function cutoff $\widetilde{\Lambda}=800 \,
{\rm MeV}$).\label{fig:phases}}
\end{center}
\end{figure*}

Upon Fourier transformation, these LO and NLO contact interactions
lead to local smeared-out delta functions $\delta_{R_0}({\bf r})$ and
their derivatives when a local regulator $f_{\rm local}(q^2)$ is
used. We implicitly define the local regulator by
taking $\delta_{R_0}({\bf r}) \sim e^{-(r/R_0)^4}$ with an exponential
regulator (with the same scale $R_0$) similarly as for the long-range
parts. We thus have for the LO contact interactions in coordinate space
\begin{equation}
\int \frac{d{\bf q}}{(2\pi)^3} \, C_{S,T} \, f_{\rm local}(q^2) \, 
e^{i{\bf q}\cdot{\bf r}} = C_{S,T} \, 
\frac{e^{-(r/R_0)^4}}{\pi \Gamma\bigl(\frac{3}{4}\bigr) R_0^3} \,,
\label{eq:LOshort}
\end{equation}
where the denominator is determined by normalization. The analogous
local expressions involving the NLO contact interactions are obtained
by replacing $C_{S,T}$ with the seven different operators of
Eq.~(\ref{eq:NLOshort}). Finally, for the range of the scale $R_0$ we
consider $R_0=0.8-1.2 \, {\rm fm}$ corresponding to typical momentum
cutoffs $\Lambda \sim 600-400 \, {\rm MeV}$ in chiral EFT interactions.
This follows Weinberg's power counting with typical cutoffs
of order the breakdown 
scale $\sim 500\, {\rm MeV}$~\cite{Lepage:1997cs,Epelbaum:2009a}.
The same local rearrangement can be applied to modified power 
counting~\cite{Nogga:2005}, to pionless 
EFT~\cite{Kirscher:2010}, to power counting that includes 
$k_{\rm F}$ as an explicit scale~\cite{Lacour:2011}, and when making use of 
off-shell ambiguities~\cite{Oset:2012}.

The low-energy couplings $C_{S,T}$ at LO plus $C_{1-7}$ at NLO and
N$^2$LO are fit in Ref.~\cite{Freunek:2007} for different $R_0$ to the
NN phase shifts of the Nijmegen partial-wave analysis~\cite{Stoks:1993}
at laboratory energies $E_{\rm lab} = 1, 5, 10, 25, 50,$ and $100 \,
{\rm MeV}$. The reproduction of the isospin $T=1$ S- and P-waves is
shown order by order in Fig.~\ref{fig:phases}, where the bands are
obtained by varying $R_0$ between $0.8-1.2 \, {\rm fm}$ and provide
a measure of the theoretical uncertainty. For the $R_0=1.2 \, {\rm fm}$
N$^2$LO NN potential, we list the low-energy couplings at LO, NLO, and
N$^2$LO in Table~\ref{tab:contacts}. At N$^2$LO, an 
isospin-symmetry-breaking contact interaction ($C_{nn}$ for neutrons)
is added in the spin $S=0$ channel (to $C_S - 3 C_T$), which is fit to
a scattering length of $-18.8 \, {\rm fm}$.  As shown in
Fig.~\ref{fig:phases}, the comparison with NN phase shifts is very
good for $E_{\rm lab} \lesssim 150 \, {\rm MeV}$. This is similar for higher partial waves and isospin $T=0$ channels, which will
be reported in a later paper that will also study improved fits. 
In cases where there are deviations
for higher energies (such as in the $^3$P$_2$ channel of
Fig.~\ref{fig:phases}), the width of the band signals significant
theoretical uncertainties due to the chiral EFT truncation at
N$^2$LO. The NLO and N$^2$LO bands nicely overlap (as shown for the
cases in Fig.~\ref{fig:phases}), or are very close, but it is also
apparent that the N$^2$LO bands are of a similar size as at NLO. This
is because the width of the bands at both NLO and N$^2$LO shows
effects of the neglected order-$Q^4$ contact interactions.

Finally, we emphasize that the newly introduced local chiral
EFT potentials include the same physics as the momentum-space versions.
This is especially clear when antisymmetrizing. Besides the new
idea of removing the $k^2$ terms, there
are no conceptual differences between the two ways of regularizing 
(see also the early work ~\cite{Ordonez:1996}).

\begin{figure}[t]
\begin{center}
\includegraphics[width=0.9\columnwidth,clip=]{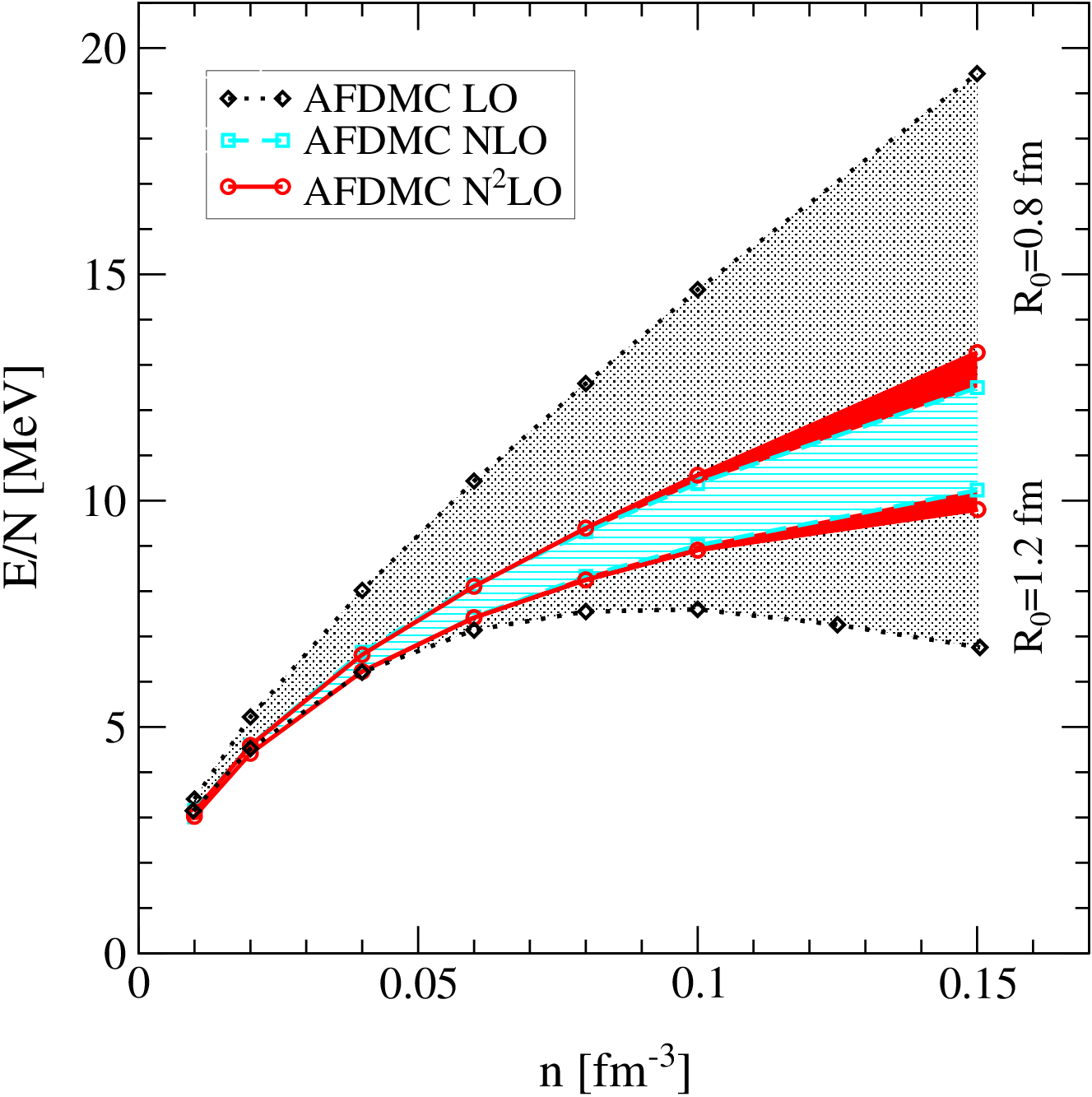}
\caption{(Color online) Neutron matter energy per particle $E/N$
as a function of density $n$ calculated
using AFDMC with chiral EFT NN interactions at LO, NLO, and N$^2$LO.
The statistical errors are smaller than the points shown.
The lines give the range of the energy band obtained by varying
$R_0$ between $0.8-1.2 \, {\rm fm}$ (as for the phase shifts in
Fig.~\ref{fig:phases}), which provides an estimate of the theoretical
uncertainty at each order. The N$^2$LO band is comparable to
the one at NLO due to the large $c_i$ couplings in the N$^2$LO
two-pion exchange.\label{fig:eos_qmc}}
\end{center}
\end{figure}

We then apply the developed local LO, NLO, and N$^2$LO chiral EFT
interactions in systematic QMC calculations for the first time. Since
nuclear forces contain quadratic spin, isospin, and tensor operators
(of the form ${\bm \sigma}^{\alpha}_i \, A^{\alpha \beta}_{ij} \, {\bm \sigma}^{\beta}_j$),
the many-body wave function cannot be expressed as a product of
single-particle spin-isospin states. All possible spin-isospin
nucleon-pair states need to be explicitly accounted for, leading to an
exponential increase in the number of possible states. As a result,
Green's Function Monte Carlo (GFMC) calculations are presently limited
to 12 nucleons and 16 neutrons~\cite{Pieper:2008}. In this Letter, we
would like to simulate $O(100)$ neutrons to access the thermodynamic
limit. We therefore turn to the auxiliary-field diffusion Monte Carlo (AFDMC) method~\cite{Schmidt:1999},
which is capable of efficiently handling spin-dependent Hamiltonians.

Schematically, AFDMC rewrites the Green's function by applying a
Hubbard-Stratonovich transformation using auxiliary fields to change
the quadratic spin-isospin operator dependences to linear. As a
result, when applied to a wave function that is a product of
single-particle spin-isospin states, the new propagator independently
rotates the spin of every single nucleon. Using this approach, central
and tensor interactions can be fully included in an AFDMC stochastic
simulation. For the case of neutrons, it has also been possible to
include fully in AFDMC spin-orbit interactions and three-body forces~\cite{Sarsa:2003,Gandolfi:2009}.

\begin{figure}[t]
\begin{center}
\includegraphics[width=0.9\columnwidth,clip=]{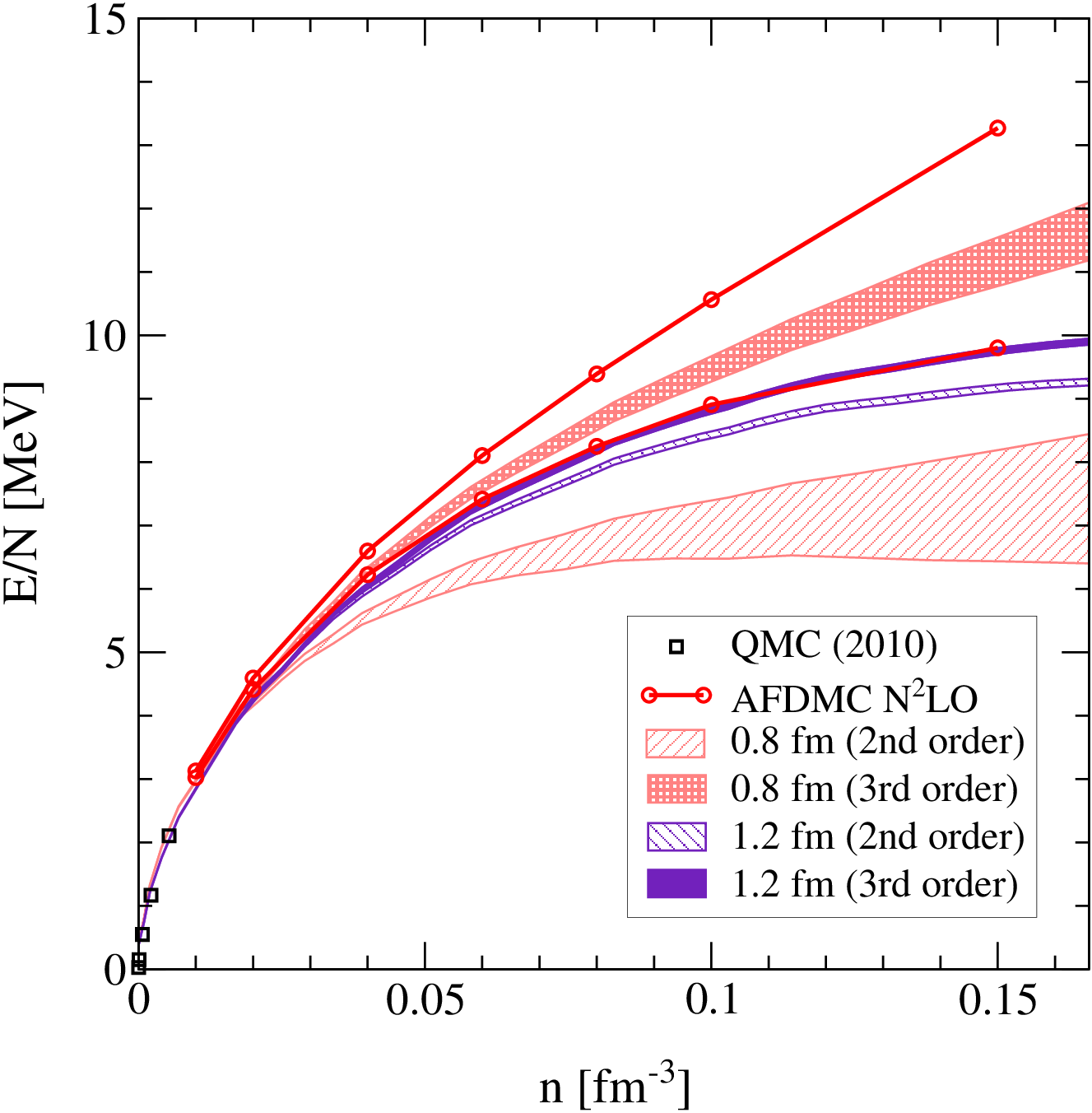}
\caption{(Color online) The AFDMC N$^2$LO band of Fig.~\ref{fig:eos_qmc}
in comparison to perturbative calculations of the neutron matter
energy using the same local N$^2$LO NN interactions. The lower (upper)
limit of the AFDMC N$^2$LO band is for $R_0 = 1.2 \, {\rm fm}$ ($R_0 =
0.8 \, {\rm fm}$), corresponding to a momentum cutoff $\Lambda \sim
400 \, {\rm MeV}$ ($\Lambda \sim 600 \, {\rm MeV}$). Perturbative
results are shown for Hartree-Fock plus second-order contributions
(2nd order) and including third-order particle-particle and hole-hole
corrections (3rd order). The bands at 2nd and 3rd order are obtained
by using a Hartree-Fock or free single-particle spectrum. For the
softer $R_0 = 1.2 \, {\rm fm}$ interaction (narrow purple bands), the
third-order corrections are small and the perturbative third-order
energy is in excellent agreement with the AFDMC results, while for the
harder $R_0 = 0.8 \, {\rm fm}$ interaction (light red bands), the
convergence is clearly slow. At low densities, we also show the QMC
(2010) results of Ref.~\cite{Gezerlis:2010,Gezerlis:2012}.\label{fig:eos_mbpt}}
\end{center}
\end{figure}

We first studied finite-size effects and the dependence on the Jastrow
correlations in the trial Jastrow-Slater wave function (in continuum
QMC calculations there are no discretization effects). The dependence on
particle number was found to be nearly identical to that of the
noninteracting Fermi system, consistent with results using
phenomenological NN potentials~\cite{Gandolfi:2009}. Therefore, we
performed calculations for an optimal number of 66 particles, while
also including contributions from the 26 cells neighboring the primary
simulation box. We also compared the neutron matter energy at a
density $0.1 \, {\rm fm}^{-3}$ starting from no to full Jastrow correlations
based on the same $R_0$ local chiral NN interactions versus Jastrow
correlations of the hard Argonne $v_8'$ potential, as a first step in
probing the general dependence on the Jastrow term. For the softer
$R_0 = 1.2 \, {\rm fm}$ ($R_0 = 0.8 \, {\rm fm}$) interactions the
changes of the energy per particle are at most $0.1 \, {\rm MeV}$
($0.6 \, {\rm MeV}$), which corresponds to $1 \%$ ($5 \%$)
changes. This appears to be related to the way the propagator is
sampled with tensor and spin-orbit interactions and will be studied in
detail in a forthcoming paper. The exact results should be independent
of the trial wave function, but we consider Jastrow correlations based
on the same $R_0$ interactions more consistent and use these.

In Fig.~\ref{fig:eos_qmc} we present first AFDMC calculations for the
neutron matter energy with chiral EFT NN interactions at LO, NLO, and
N$^2$LO. Our results represent nonperturbative energies for
neutron matter based on chiral EFT beyond low densities.  For
neutrons, the AFDMC method has been carefully benchmarked with nuclear
GFMC, 
which can handle beyond-central correlations as well as 
release the nodal or phase constraint after convergence to the ground state. 
Both have been found to have minimal effects on the equation of 
state of neutrons~\cite{Gandolfi:2009,Carlson:2003,Gandolfi:2011}.
At each order, the full interaction is
used both in the propagator and when evaluating observables.
The bands in Fig.~\ref{fig:eos_qmc} give the range of the energy
obtained by varying $R_0$ between $0.8-1.2 \, {\rm fm}$, where the
softer $R_0 = 1.2 \, {\rm fm}$ interactions yield the lower
energies. At low densities (low Fermi momenta), as expected the energy
is well constrained at LO, with small corrections at NLO due to
effective range effects~\cite{dEFT,Gezerlis:2010}. AFDMC enables us to
present results up to saturation density (and higher, but we emphasize
that the contributions of 3N forces will become important for
densities $n \gtrsim 0.05 \, {\rm fm}^{-3}$~\cite{Hebeler:2010a}). At
LO, the energy has a large uncertainty. The overlap of the bands
at different orders in Fig.~\ref{fig:eos_qmc} is excellent. In
addition, the result that the NLO and N$^2$LO bands are comparable is
expected due to the large $c_i$ entering at N$^2$LO; this is similar
to the phase shift bands in Fig.~\ref{fig:phases}. 
At the 
highest density studied, the 
size of the N$^2$LO band is approximately 10\% of the
potential energy, which will be improved by including 3N forces~\cite{Hebeler:2010a} and
going to higher order~\cite{Tews:2013}.
Therefore, our
first QMC results for neutron matter exhibit a systematic
order-by-order convergence in chiral EFT. 
Given the small contributions coming from 3N forces at intermediate
density, as well as the limited size of the systematic error bands
there, our results are a nonperturbative benchmark that can lead to
further predictions at higher density, when 3N forces are consistently
included.

Our AFDMC results provide first nonperturbative benchmarks for chiral
EFT interactions at nuclear densities. We have performed perturbative
calculations following Refs.~\cite{Hebeler:2010a,nucmatt,Tews:2013}
based on the same local N$^2$LO NN interactions at the Hartree-Fock
level plus second-order contributions and including third-order
particle-particle and hole-hole corrections. At each order, we give
bands obtained by using a Hartree-Fock or free single-particle
spectrum. The perturbative energies are compared in
Fig.~\ref{fig:eos_mbpt} to the AFDMC N$^2$LO results.  For the softer
$R_0 = 1.2 \, {\rm fm}$ ($\Lambda \sim 400 \, {\rm MeV}$) interaction,
the third-order corrections are small and the perturbative third-order
energy is in excellent agreement with the AFDMC results, while for the
harder $R_0 = 0.8 \, {\rm fm}$ ($\Lambda \sim 600 \, {\rm MeV}$)
interaction, the convergence is clearly slow. This is the first
nonperturbative validation for neutron matter of the possible perturbativeness
of low-cutoff $\Lambda \sim 400 \, {\rm MeV}$ interactions~\cite{PPNP}.
Finally, in the low-density regime, the results in Fig.~\ref{fig:eos_mbpt}
match the QMC calculations of Ref.~\cite{Gezerlis:2010,Gezerlis:2012}
based on central interactions that reproduce the large neutron-neutron
scattering length and the effective range physics.

In summary, we have presented the first QMC calculations with chiral
EFT interactions. This was achieved by using a freedom in chiral EFT
to remove all sources of nonlocality to N$^2$LO. We have constructed
local LO, NLO, and N$^2$LO NN interactions, given in operator form
times local potentials $V(r)$ in coordinate space. The reproduction of
the NN phase shifts is very good compared to the momentum-space
N$^2$LO NN potentials of~Ref.~\cite{EGMN2LO}. Direct application of
the local chiral NN interactions in AFDMC sets first nonperturbative
benchmarks for the neutron matter equation of state at nuclear
densities. Our results show systematic order-by-order convergence with
theoretical uncertainties and validate perturbative calculations for
the softer local NN interactions. Future AFDMC calculations with local
N$^2$LO 3N forces will provide {\it ab initio} constraints for nuclear
density functionals and for dense matter in astrophysics. This work
paves the way for QMC calculations with systematic chiral EFT
interactions for nuclei, neutron drops, and nuclear matter. 
Regarding nuclear matter, a perturbative approach has been able to
predict realistic saturation properties using parameters 
fit only to few-body systems~\cite{nucmatt}, so future QMC work will
be key to validating this and to providing nonperturbative benchmarks.
By direct matching to lattice QCD results~\cite{latticeQCD} 
(for example, for few-neutron systems in a box) also varying the pion 
mass in chiral EFT, the approach presented here will be able to 
connect nuclear physics to the underlying theory of QCD.

\begin{acknowledgments}
We thank M.~Freunek for the NN interaction fits performed in Ref.~\cite{Freunek:2007} and J.~Carlson, T.~Kr\"uger, J.~Lynn, and
K.~Schmidt for stimulating discussions. This work was carried out
within the ERC Grant No.~307986 STRONGINT, and also supported by the
Helmholtz Alliance Program of the Helmholtz Association contract
HA216/EMMI ``Extremes of Density and Temperature: Cosmic Matter in the
Laboratory'', the ERC Grant No.~259218 NuclearEFT, the US DOE SciDAC-3
NUCLEI project, the LANL LDRD program, and by the NSF under Grant 
No.~PHY--1002478. Computations were performed at the J\"{u}lich 
Supercomputing Center and at NERSC.
\end{acknowledgments}

\end{document}